\documentclass[aps,prl,floatfix,twocolumn,footinbib,superscriptaddress]{revtex4}
\usepackage{amsmath}
\usepackage{epsfig}
\usepackage{amsfonts}
\usepackage{amssymb}
\usepackage{graphicx}
\usepackage{epstopdf}
\begin{document}
\title{The magneto-optical Faraday effect in spin liquid candidates}
\author{Jacob R. Colbert}
\affiliation{Department of Physics, Massachusetts Institute of Technology, Cambridge, MA 02139, USA}
\author{H. Dennis Drew}
\affiliation{Department of Physics, University of Maryland at College Park, College Park, MD 20742, USA}
\affiliation{Center for Nanophysics and Advanced Materials, University of Maryland at College park, College Park, Maryland, 20742, USA}
\author{Patrick A. Lee}
\affiliation{Department of Physics, Massachusetts Institute of Technology, Cambridge, MA 02139, USA}
\begin{abstract}
We propose an experiment to use the magneto-optical Faraday effect to probe the dynamic Hall conductivity of spin liquid candidates. Theory predicts that an external magnetic field will generate an internal gauge field. If the source of conductivity is in spinons with a Fermi surface, a finite Faraday rotation angle is expected. We predict the angle to scale as the square of the frequency rather than display the standard cyclotron resonance pattern. Furthermore, the Faraday effect should be able to distinguish the ground state of the spin liquid, as we predict no rotation for massless Dirac spinons.  We give a semiquantitative estimate for the magnitude of the effect and find that it should be experimentally feasible to detect in both $\kappa$-(ET)$_2$Cu$_2$(CN)$_3$ and, if the spinons form a Fermi surface, Herbertsmithite. We also comment on the magneto-optical Kerr effect and show that the imaginary part of the Kerr angle may be measurable.
\end{abstract}
\maketitle
\section{Introduction}
Recent experiments in the spin liquid candidates Herbertsmithite and $\kappa$-(ET)$_2$Cu$_2$(CN)$_3$  have observed a power law in the conductivity below the Mott gap \cite{Pilon-2013,Elasser-2012}. One of the potential explanations of this conductivity is optical excitations of spinons in a spin liquid state\cite{Potter-2013,Ng-2007}. In the experiments on Herbertsmithite, the measured conductivity amplitude and exponent are slightly smaller but comparable to the theoretical predictions for the spinon contribution to the conductivity. Here we propose an experiment using the magneto-optical rotation of light as a further probe of possible contributions of spinons to the finite frequency conductivity tensor.

The organic material $\kappa$-(ET)$_2$Cu$_2$(CN)$_3$ is believed to be a spin liquid with spinon excitations forming a Fermi surface\cite{Lee-2005,Yamashita-2008}. While it is commonly believed that Herbertsmithite has a spin liquid phase, it is unclear what the ground state is. Projected wave function studies predict a spin liquid state with massless Dirac fermions\cite{Ran-2007}, while density matrix renormalization group calculations find a gapped $\mathbb{Z}_2$ spin liquid\cite{Yan-2011}. Neutron scattering and thermodynamic measurements show evidence of gapless excitations\cite{Han-2012,Han-2014}. The neutron scattering pattern shows gapless spin excitations across a wide range of momentum transfer, $\mathbf{Q}$, potentially suggesting that the spinons form a Fermi surface rather than there only being two Dirac nodes where the excitations are gapless. In addition, the heat capacity showed a linear T term in high magnetic field \cite{Han-2014}. We show below that massless Dirac spinons should show no linear magneto-optical Faraday effect, while the Faraday rotation should be experimentally observable for spinons with a Fermi surface, allowing an experimental probe to distinguish between the two gapless ground states. 

As was shown by Motrunich and others, in the presence of the magnetic field the spinons will see an internal magnetic field due to a linear coupling between the physical magnetic field and the gauge magnetic field.\cite{Motrunich-2006} This breaking of time reversal symmetry for the spinons should be observable through the measurement of the rotation of the polarization of the transmitted light. The Faraday rotation at normal incidence is given, for small rotations, by
\begin{equation}
\theta_F=\frac{\ell}{nc}2\pi \sigma_{xy}'(\mathrm{3D})
\label{theta_F}
\end{equation}
where $\ell$ is the thickness in the direction of propagation of the light, $n$ is the index of refraction, and $\sigma_{xy}'(\mathrm{3D})$ is the real part of the off-diagonal in-plane 3D conductivity.
\section{Theoretical Background}
Herbertsmithite is a  Mott insulator and can be well described by taking a strong coupling $t/U$ expansion of the half-filled Hubbard model. While $\kappa$-(ET)$_2$Cu$_2$(CN)$_3$ is just on the insulating side of the Mott transition, we assume that a similar expansion is still an appropriate starting point. In the limit where the electron hopping term vanishes, $t=0$, the ground state is given by single occupation of each lattice site and has a $2^N$-fold degeneracy. As we increase the hopping relative to the Coulomb energy, $U$, the degeneracy is broken and the ground state is lowered by mixing of the different singly occupied states through virtual hopping. We can project the Hamiltonian onto the low energy manifold, to get an effective spin Hamiltonian for the system. To lowest order in $t/U$ we get a Heisenberg antiferromagnetic interaction with $J=4t^2/U$. Higher order terms will introduce further spin-spin interactions, as well as loop interactions that will be important when we compute the coupling between the physical and emergent magnetic fields. 

The spin model can be solved approximately by introducing fermions to carry the spin in an enlarged Hilbert space replacing each spin with $\mathbf{S_i}=f^\dagger_{i,a}\boldsymbol{\sigma}_{ab}f_b$. The physical Hilbert space is the subspace with each state singly occupied. After making this substitution into the spin Hamiltonian and introducing an integration over an auxillary scalar field to enforce the constraint \cite{Wen-2004}, we apply a mean field treatment of the resulting 4-Fermi term to get the mean field Hamiltonian
\begin{equation}
H_{MF}=J\sum_{\langle ij \rangle}\left(\chi_{ij}f^\dagger_{i,\sigma}f_{j,\sigma}+c.c.\right)-\lambda\sum_{i,\sigma}f^\dagger_{i,\sigma}f_{i,\sigma}
\end{equation}
with $\chi_{ij}=\langle f^\dagger_if_j\rangle$ and $\lambda$ enforcing the constraint of one fermion per site on average. We can interpret $f^\dagger$ as the creation operator of fermionic spinons. To get back to a physical spin wavefunction, the solution of this mean field Hamiltonian can be projected onto the physical single occupancy subspace. Allowing fluctuations of $\lambda$ and the phase of the hopping term $\chi_{ij}$ corresponds to introducing a dynamical electric and magnetic vector potential. This is the source of the emergent electic and magnetic fields, $\mathbf{e}$ and $\mathbf{b}$, that the spinons feel.
\subsection{Mechanism for optical conductivity from spinons}
In order to compute the conductivity we follow the framework given by Potter \textit{et al}\cite{Potter-2013}. The physical conductivity is proportional to the correlation function of the emergent gauge electric field
\begin{equation}
\sigma_{ij}\approx 72\pi(n_\Delta a^2)\frac{e^2}{h}i\omega\frac{t^2}{U^4}\langle e_\omega^i e_{-\omega}^j\rangle
\label{eq:first}
\end{equation}
where $n_\Delta$ is the density of triangles in the lattice, $a$ is the lattice constant, and $\mathbf{e}=\boldsymbol{\nabla}\lambda+\dot{\mathbf{a}}$ is the gauge electric field with $\mathbf{a}$ the continuum vector potential corresponding to the phase of $\chi_{ij}$. They compute this correlation function within the random-phase approximation and find that 
\begin{equation}
\langle e_\omega^i e_{-\omega}^j\rangle=-\frac{i\omega}{2}[\sigma_s^{-1}]_{ij}=-\frac{i\omega}{2}[\rho_s]_{ij}
\end{equation}
where $\rho_s$ is the spinon resistivity in response to the internal gauge electric field. Substituting this into equation \ref{eq:first} we get, 
\begin{equation}
\sigma_{ij}\approx 36\pi(n_\Delta a^2)\frac{e^2}{h}i\omega\frac{t^2}{U^4}[\rho_s]_{ij}
\label{eq:potter}
\end{equation}
This equation is valid only for frequencies less than the spinon bandwidth, which is estimated to be on the order of $J$. For the materials discussed here $J\approx 250-350$ K, corresponding to frequencies of about $5-7$ THz.

\subsection{Coupling between the physical and emergent magnetic fields}
We must first calculate the magnitude of the induced internal flux that the spinons feel. This was done already for the organic material $\kappa$-(ET)$_2$Cu$_2$(CN)$_3$ by Motrunich using a strong coupling expansion. A perturbative $t/U$ expansion of the Hubbard model leads to a linear coupling of the applied magnetic field to the scalar spin chirality, $\mathbf{S_1}\times\mathbf{S_2}\cdot\mathbf{S_3}$. Bulaevskii \emph{et al}. showed that virtual charge fluctuations lead to a current (and orbital magnetic moment) proportional to the spin chirality\cite{Bulaevskii-2008}. The coupling shown by Motrunich can be more physically interpreted as the coupling of the external magnetic field and this orbital magnetic moment. This spin chirality can be interpreted additionally as a Berry's flux or the emergent magnetic field that the charge neutral spinons feel\cite{Wen-1989,Lee-1992}.

In order to estimate the magnitude of the internal field produced, Motrunich minimized the energy as a function of the applied field. In the organic, the linear coupling found at third order by hopping an electron around a triangle is supplemented by a fourth order term that is quadratic in the internal flux, $\Phi_\mathrm{int}$, that stabilizes the field giving a mean field energy per site in the uniform flux state of 
\begin{equation}
E_{\mathrm{mf}}=\alpha \Phi_{\mathrm{ext}}\frac{t^3}{U^2}\sin(\Phi_{\mathrm{int}})+\beta \frac{t^4}{U^3}\cos(2\Phi_{\mathrm{int}})
\end{equation}
keeping only the most relevant terms. $\alpha$ and $\beta$ contain numerical coefficients and material dependent parameters and $\Phi_{\mathrm{ext}}$ and $\Phi_{\mathrm{int}}$ the dimensionless fluxes per triangle. The factor of $2$ on the flux in the second term comes from the fact that the fourth order loop encloses two triangles. Balancing these terms, along with other corrections, he found that the emergent flux was related to the external flux by 
$\Phi_{int}=\Gamma \Phi_{ext}$, with $\Gamma\sim 1\mathrm{-}2$.\cite{Motrunich-2006} We will use a value of $\Gamma=1.5$ for the rest of the paper.%

\begin{figure}
\includegraphics[scale=.8]{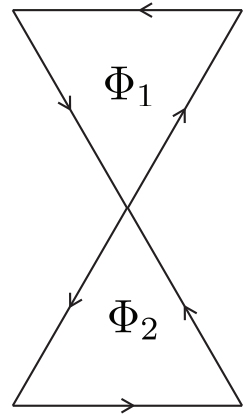}
\caption{Illustration of the virtual hopping needed to give the restoring force on the Kagom\'{e} lattice.}
\label{fig:kagome}
\end{figure}

We expect the Kagom\'{e} lattice of Herbertsmithite to give a similar result. The argument is similar, but one needs to go to a higher order in $t/U$ to get a restoring term quadratic in $b$. It is not until order $t^6/U^5$ that a term that is even in the emergent flux arises, given by hopping around two corner sharing triangles (see figure \ref{fig:kagome}), i.e. the mean field energy per site is given instead by 
\begin{equation}
E_{\mathrm{mf}}=\alpha \Phi_{\mathrm{ext}}\frac{t^3}{U^2}\sin(\Phi_{\mathrm{int}})+\beta \frac{t^6}{U^5}\cos(2\Phi_{\mathrm{int}})
\end{equation}
This effect would tend to make the emergent field stronger, as it decreases the effective gauge stiffness. This is counteracted, however, by a larger combinatorial prefactor for the term $\beta$.

The fact that as one goes further into the insulating phase, with $t/U$ getting smaller, the emergent magnetic field grows is counterintuitive and points to a potential limitation of the theory. As $t/U$ shrinks the effective gauge stiffness does as well, meaning that the system will be more prone to gauge fluctuations. This calls into question the accuracy of our expansion, as we neglect screening affects. However, since even for Herbertsmithite we are only in the intermediate regime, we expect that this treatment will suffice to get an estimate of the effect, and we use $\Gamma=1.5$ for Herbertsmithite as well.

\section{Semi-quantitative estimates of the Faraday Effect}
We can now calculate the spinon conductivity within each of the possible ground states. For a massless Dirac spin liquid in the presence of a static gauge magnetic field, we expect the spinon bands to be Zeeman split, creating a hole pocket of spin-down spinons and a spinon pocket of spin-up spinons. The Hall effect due to the particle pocket will be canceled by that due to the hole pocket, and we expect no Hall conductivity and thus no Faraday rotation that is linear in $H$. This is in contrast to the case of a spinon Fermi surface which, as we now show, should have an experimentally detectable Faraday effect, allowing experiments to distinguish between these two gapless spin liquid phases.

For a spinon Fermi surface we expect to see an effect. Within experimentally realizable fields, the Landau level filling factor should be very large, and we model the spinon conductivity with the Drude model. Within the Drude model, we get that the spinon resistivity is given by 	
\begin{equation}
\rho_s=\frac{m_s}{n}\left(\begin{array}{cc}
\gamma-i\omega & \omega_c \\
-\omega_c       & \gamma-i\omega 
\end{array}\right)
\label{eq:rho_s}
\end{equation}
where $m_s$ is the spinon mass, $n$ is the spinon density, $\gamma$ is the spinon scattering rate, and $\omega_c$ is the spinon cyclotron frequency. The spinon bandwidth is estimated to be a fraction of $J$, which corresponds to a spinon mass, $m_s\sim 1/(Ja^2)$. The spinon cyclotron frequency is given by $\omega_c=b/m_s=\Gamma B/m_s$.

Putting all this together we get
\begin{align}
\sigma_{xy}&\sim 36\pi(n_\Delta a^2)(\frac{t}{U})^2(\frac{\omega}{U})^2 \Gamma \frac{B}{n}\frac{e^2}{h}\\
           &=72\pi^2 \frac{n_\Delta}{n}(\frac{t}{U})^2(\frac{\omega}{U})^2 \Gamma \frac{Ba^2}{\phi_0}\frac{e^2}{h}
\end{align}
where $\phi_0$ is the magnetic flux quantum.

 We can now estimate the magnitude of this conductivity and of the Faraday rotation in both $\kappa$-(ET)$_2$Cu$_2$(CN)$_3$ and Herbertsmithite. In the organic, we use $t=55$ meV and $t/U=.12$\cite{Motrunich-2006}. We use a lattice constant of $a=10$ \r{A} and an interlayer spacing of $d=16$ \r{A}\cite{Katsura-2010}. There are two triangles per spinon, so $n_\Delta/n=2$. We calculate the Hall conductivity of one layer $\sigma_{xy}'\sim 4\cdot 10^{-6} \frac{e^2}{h}$ at $\omega=2\pi\times1$ THz and $7$ T. This in turn gives a 3D conductivity $\sigma_{xy}(3\mathrm{D})\sim 1\cdot10^{-3}\mathrm{ }\Omega^{-1}\mathrm{cm}^{-1}$. Using equation \ref{theta_F}, we find an estimate of the Faraday rotation to be about $0.2$ mrad for a $30$ $\mu$m sample using a dielectric constant of $\sim 4$ \cite{Nakamura-2009}. Extrapolating from IR data, we estimate that there should be reasonable transmission through this thickness at $1$ THz \cite{Elasser-2012}. Recent experiments have resolutions of down to 30 $\mu\mathrm{rad}$, so this effect should be within experimental limitations \cite{Jenkins-2012}. The rotation angle, like the Hall conductivity, should be quadratic in the frequency and linear in the magnetic field.

In Herbertsmithite, assuming a spinon Fermi liquid ground state, we use $t=100$ meV and $t/U\sim.1$, a lattice constant $a\sim 10$ \r{A}, and an interlayer spacing of $d\sim 10$ \r{A} \cite{Potter-2013}. On the Kagom\'{e} lattice, $n_\Delta/n=2/3$. The single layer Hall conductivity is smaller by an order of magnitude resulting in a single layer conductivity of $\sigma^{xy}\sim 2\times 10^{-7} \frac{e^2}{h}$, a 3d conductivity of $\sigma_{xy}(3\mathrm{D})\sim4\cdot10^{-5}\mathrm{ }\Omega^{-1}\mathrm{cm}^{-1}$, and a Faraday rotation of $0.2$ mrad at $\omega=2\pi\times1$ THz and $7$ T in a $0.3$ mm thick sample. This rotation should again still be observable with the current resolution of experimental setups.
 
We note that the predicted frequency dependence of the Faraday rotation due to spinons is distinct from that due to conductivity from electronic sources.  Typically, in low carrier density metals, when the  Hall conductivity is from electronic sources, the Faraday rotation angle shows a resonance structure around the electron cyclotron frequency with an $(\omega-\omega_c)^{-1}$ tail for frequencies away from the resonance. On the other hand, the resistivity tensor does not show a resonance, as seen in equation \ref{eq:rho_s}. In particular, $\rho_{xx}$ does not depend on the magnetic field. However in our case, for conductivity due to spinons, there is no resonance peak in the conductivity or Faraday angle because the physical conductivity is proportional to the spinon resistivity tensor. Instead, we expect the Hall conductivity to scale as  $\omega^2$ and for there to be no magneto-conductance. 

If spinons are the dominant source of conductivity, it is the physical resistivity that shows a resonance. If an experiment could accurately measure both $\sigma_{xx}$ and $\sigma_{xy}$, then, by inverting the conductivity tensor to get the resistivity tensor, we expect to see a resonance at the spinon cyclotron frequency. This would give direct evidence of the presence of both spinons and the emergent gauge field. We expect the cyclotron frequency to be about $\omega_c\sim 2\pi\times 50$ GHz. However, the expected Faraday rotation at this frequency is less than $5\cdot 10^{-6}$ rad, in both materials, which is below the resolution of current experiments.

We can also take a look at the longitudinal conductivity contribution from spinons using the Drude model. Using equation \ref{eq:potter} we get, 
\begin{align}
\sigma_{xx}&\sim 36\pi(n_\Delta a^2)(\frac{t}{U})^2(\frac{\omega}{U})^2 \frac{m_s(\gamma-i\omega)}{n}\frac{e^2}{h}\\
		   &\sim 36\pi\frac{n_\Delta}{n}(\frac{t}{U})^2(\frac{\omega}{U})^2\frac{\gamma-i\omega}{J}\frac{e^2}{h}
\end{align}
Assuming that the scattering rate is dominated by inelastic scattering in our temperature range, we take the factor $\gamma\sim k_B T\sim J/10 $. We also estimate $m_s\sim 1/(J a^2)$. This leads to a quadratic power law for the real part of the conductivity. For Herbertsmithite, our crude estimate predicts a value $\sigma_{xx}\sim 1\cdot 10^{-6}\frac{e^2}{h}$ at $\omega=2\pi\cdot 1$ THz, which is a couple orders of magnitude smaller than the conductivity observed by Pilon \textit{et al}. \cite{Pilon-2013}. On the other hand, the Dirac spin liquid model gave a reasonable estimate \cite{Potter-2013}. In both models, we expect that the longitudinal conductivity should show no field dependence, as observed in Herbertsmithite \cite{Pilon-2013}. 

We comment also on the magneto-optical Kerr effect, the rotation of the polarization of the reflected light. For normal incidence the complex Kerr angle is given by
\begin{equation}
\theta_K=i\left(\frac{n_+-n_-}{n_+n_--1}\right)
\end{equation}
where $n_\pm$ are the indices  of refraction for right and left circularly polarized light\cite{Argyres-1955}. The imaginary part of the Kerr angle, $\theta_K''$, gives the ellipticity of the polarization of the light reflected from incident linearly polarized light, measured as the ratio of the major to minor axes. The real part, $\theta_K'$, gives the rotation angle between the initial polarization and the major axis of the final polarization. Because our predicted Hall conductivity is real, we have to expand this expression to second order in the conductivity to get a non-vanishing contribution to the rotation angle. We find that 
\begin{equation}
\theta_K'\approx \left(\frac{32}{(\epsilon-1)^2}-\frac{8}{(\epsilon-1)\epsilon}\right)\frac{\pi^2}{\omega^2\sqrt{\epsilon}}\sigma_{xx}'\sigma_{xy}'
\end{equation}
with $\epsilon$ the dielectric constant. For these materials at realizable fields the rotation is beyond the resolution of current instruments. For the organic, we predict $\theta_K'\sim 10^{-8}$ rad. One potential way to boost this value is to tune the frequency to a phonon resonace in order to boost the value of the diagonal conductivity, while the off-diagonal part should be unaffected and still due only to the spinon contribution. 

However, the ellipticity of the reflected light should be observable. The imaginary part of the Kerr angle only requires approximating to first order in conductivity, 
\begin{equation}
\theta_K''=\approx \frac{4\pi}{n(n^2-1)\omega}\sigma_{xy}'
\end{equation}
Within the spinon Fermi surface model, we predict that the organic will have an ellipticity $\theta_K''\approx 2$ mrad and Herbertsmithite will have $\theta_K''\approx 0.1$ mrad. This ellipticity is directly measurable and is within experimental limitations. We expect the ellipticity to be linear in both magnetic field and frequency. This effect is unexpected in an insulator. In addition, the frequency dependence is quite different from that of electrons where the ellipticity is resonant at the cyclotron frequency and falls as $(\omega(\omega-\omega_c))^{-1}$ above the resonance. Thus its observation should be a clear signature of spinon conductivity.

PAL acknowledges the support of NSF under grant DMR-1104498. HDD acknowledges the support of NSF under grant DMR-1104343.

\bibliographystyle{apsrev}
\bibliography{references}
\end{document}